\documentstyle[emulate_apj,apjfonts,epsf]{article}

\gdef\h50min{$h_{50}^{-1}$}
\gdef\1054{MS\,1054$-$03}
\gdef\kms{km\,s$^{-1}$}
\lefthead{van Dokkum et al.}
\righthead{Evolution of early-type galaxies to $z=0.83$}
\slugcomment{To be published in ApJ Letters, Volume 504, September 1, 1998}
\begin{document}
\title{Luminosity Evolution of Early-Type Galaxies to {\rm $z=0.83$}:\\
Constraints on Formation Epoch and $\Omega$}

\author{
Pieter G. van Dokkum\altaffilmark{1,2},
Marijn Franx\altaffilmark{2,1}, Daniel D. Kelson\altaffilmark{3,4},
and Garth D. Illingworth\altaffilmark{4}
}

\altaffiltext{1}
{Kapteyn Astronomical Institute, P.O. Box 800, NL-9700 AV,
Groningen, The Netherlands}
\altaffiltext{2}
{Leiden Observatory, P.O. Box 9513, NL-2300 RA, Leiden, The Netherlands}
\altaffiltext{3}
{Department of Terrestrial Magnetism, Carnegie Institution of
Washington, 5241 Broad Branch Road, NW, Washington D.C., 20015}
\altaffiltext{4}
{University of California Observatories\,/\,Lick Observatory,
University of California, Santa Cruz, CA 95064}

\begin{abstract}

We present deep spectroscopy with the Keck telescope of eight galaxies
in the  luminous X-ray cluster \1054 at $z=0.83$.  The data are combined
with imaging observations from the {\it Hubble  Space Telescope} (HST). 
The spectroscopic data are used to measure the internal kinematics of
the galaxies, and the HST data to measure their structural parameters.
Six galaxies have early-type spectra, and two have ``E+A'' spectra.
The galaxies with early-type spectra define a tight Fundamental
Plane (FP) relation.  The evolution of the mass-to-light ratio is
derived from the FP.  The $M/L$ ratio evolves as $\Delta \log M/L_B
\propto -0.40z$ ($\Omega_{\rm m}=0.3$, $\Omega_{\Lambda}=0$).
The observed evolution of the
$M/L$ ratio provides a combined constraint on the formation redshift of
the stars, the IMF, and cosmological parameters.  For a Salpeter IMF
($x=2.35$) we find that $z_{\rm form}>2.8$ and $\Omega_{\rm m}<0.86$
with 95\,\% confidence. The constraint on the
formation redshift is weaker if $\Omega_{\Lambda}>0$:
$z_{\rm form}> 1.7$ if $\Omega_{\rm m}=0.3$ and $\Omega_{\Lambda}=0.7$.
At present the limiting factor in constraining $z_{\rm form}$ and
$\Omega$ from the observed luminosity evolution of
early-type galaxies is the poor understanding of the IMF.
We find that if $\Omega_{\rm m}=1$ the IMF must be
significantly steeper than the Salpeter IMF ($x>2.6$).

\end{abstract}

\section{Introduction}

Measurements of the masses and mass-to-light ($M/L$) ratios of
early-type galaxies in distant clusters provide important constraints on
galaxy evolution.  The luminosity evolution of early-type galaxies can be
determined by comparing the luminosities of galaxies of similar masses
at different redshifts.  Also, mass measurements can be used to
constrain the merger rate. 

The evolution of the $M/L$ ratio can be measured from the evolution of
the fundamental plane of early-type galaxies.  The fundamental plane
(FP) is a tight correlation between the structural parameters and the
velocity dispersion of the form $r_{\rm e} \propto I_{\rm
e}^{-0.83}\sigma^{1.20}$ in the $B$ band
(Djorgovski \& Davis 1987; Dressler et al.\ 1987;
J\o{}rgensen, Franx \& Kj\ae{}rgaard 1996 [JFK]).
Assuming homology, the existence of the
FP implies that $M/L \propto M^{0.25}r_{\rm e}^{-0.04}$, with low
scatter (Faber et al.\ 1987).  Therefore, the evolution of the intercept
of the FP is proportional to the evolution of the mean $M/L$ ratio (see
Franx 1993a). 

Studies of the FP out to $z \approx 0.6$ have shown
that the $M/L$ ratio of massive cluster galaxies evolves slowly
(van
Dokkum \& Franx 1996 [vDF]; Kelson et al.\ 1997 [KvDFIF]; Bender et al.\
1998; Pahre 1998).  The implication is that the stars in massive
cluster galaxies were probably formed at much higher redshifts.  These
results are consistent with other studies of the evolution of the
luminosities, colors, and absorption line strengths
of bright cluster galaxies (e.g., Aragon-Salamanca et al.\
1993, Bender, Ziegler, \& Bruzual 1996,Schade, Barrientos, \& Lopez-Cruz\ 1997,
Ellis et al.\ 1997, Stanford, Eisenhardt, \& Dickinson 1998). 

The constraints on the formation epoch of massive galaxies can be made
stronger by extending the studies of the evolution of the $M/L$ ratio to
higher redshifts. In this {\em Letter}, we report on spectroscopic and
photometric observations of eight bright galaxies in the luminous X-ray
cluster \1054 at $z=0.83$.  The fundamental plane is derived, and the
evolution of the $M/L$ ratio is established out to $z=0.83$, when the
universe was 50\,\% of its present age ($q_0 = 0.15$).

\section{Spectroscopy}


\begin{figure*}[t]
\begin{center}
\leavevmode
\hbox{%
\epsfxsize=16.5cm
\epsffile{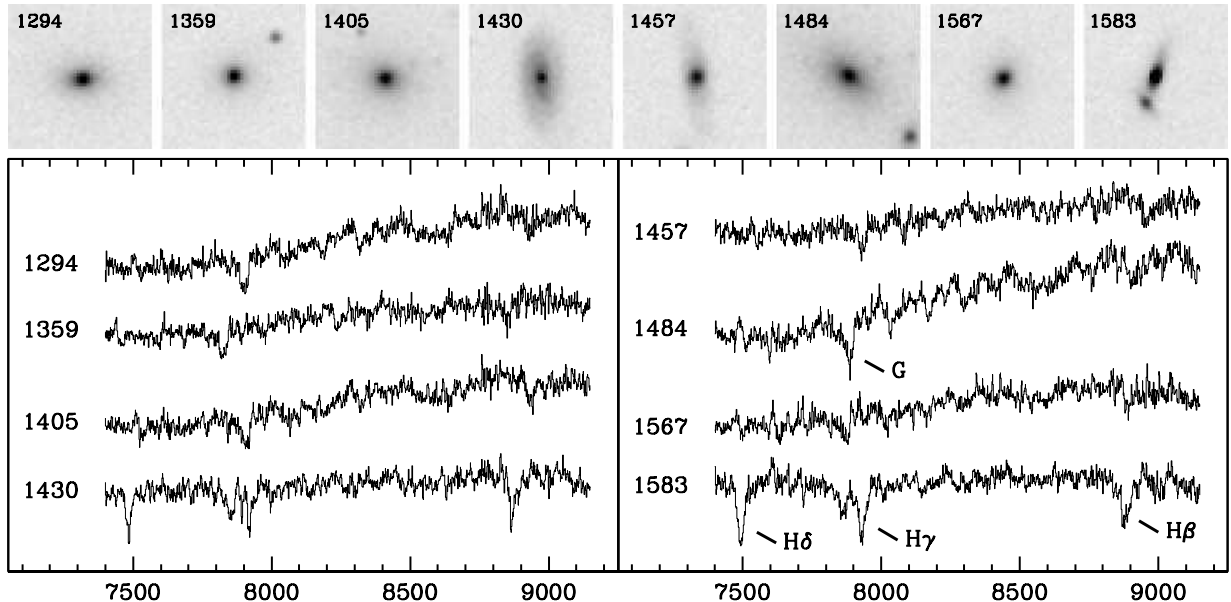}}
\figcaption{
\small HST images and Keck spectra
of the galaxies in \1054 with measured dispersions. The most prominent
spectral feature in most galaxies is the 4300\,\AA{} G band observed
at $\approx 7870$\,\AA.
Galaxies 1430 and 1583 have strong Balmer
absorption lines, and are ``E+A'' galaxies.
Galaxy 1484 is the brightest cluster galaxy.
\label{galaxies.plot}}
\end{center}
\end{figure*}

Galaxies in \1054 were selected on the basis of their $I$
band flux ($I < 22.1$),
and their location in the $B-R$ versus $R-I$ plane. 
The allowed range in color was broad, thus minimizing any
bias against passively-evolving
blue cluster galaxies. There  was no selection on morphology.
The ground based imaging data used for the selection were kindly provided
by G.\ Luppino, and are described in Luppino \& Kaiser (1997).

The cluster was observed on February 11, 1997
with LRIS (Oke et al.\ 1995) on the 10\,m W.\ M.\ Keck telescope.
The multi-aperture slitlets
were $1\farcs0$ wide, and the seeing was $0\farcs9$ --
$1\farcs0$. The instrumental resolution $\sigma_{\rm instr} \approx
50$\,\kms. The total exposure time was 14\,400\,s. A bright blue star
was included in the masks for the purpose of correcting for the H$_2$O
absorption at $7600$\,\AA{} from our atmosphere. The reduction
procedures were very similar to those described by vDF and KvDFIF. Eight
of the observed galaxies are covered by a deep HST image of the cluster
core. In this {\em Letter}, we will limit the discussion to these
galaxies.


The spectra of the galaxies are shown in Fig.\,1.
Two of the eight galaxies (1430 and 1583) have strong Balmer absorption lines
(EW$>4$\,\AA{} in the rest frame), indicating the presence of
a young stellar component.
Subsequent low resolution spectroscopy covering a larger wavelength
range showed that both galaxies have O\,{\sc ii} 3727 EW$<5$\,\AA{},
and hence are classified
``E+A'' galaxies (Dressler \& Gunn 1983).


We determined the central velocity dispersions of the galaxies from a
fit to a convolved template star in real space,
following the procedures outlined in vDF and
KvDFIF. The spectral regions that are dominated by bright sky lines were
given low weight in the fit. The Balmer lines in the E+A galaxies were
masked. The resulting dispersions, corrected to an aperture of
$3\farcs4$ at the distance of Coma (cf.\ vDF), are listed in Table 1. We
experimented with the choice of template star, the continuum filtering,
and the fitting method (e.g., fitting in Fourier space rather than real
space) to assess the systematic uncertainties, and estimate them to be
$\sim 5$\,\%.

There are several galaxies with dispersions in the range $250$
-- $300$\,\kms, showing that massive galaxies exist at $z=0.83$. The
dispersions of the two E+A galaxies are $153\pm 12$\,\kms{}
and $212\pm 25$\,\kms.
These values are significantly higher than the typical dispersions of
E+A galaxies at lower redshift ($\sim 100$\,\kms;
Franx 1993b, KvDFIF, Kelson et al.\ 1998).

\section{Photometry}


Hubble Space Telescope (HST) WFPC2 imaging data of the central parts of
\1054 were taken by Donahue et al.\ (1998) on March 13, 1996 through the
$F814W$ filter.  The total exposure time was 15\,600\,s.  
Following the usual reduction process the images were deconvolved with
the {\sc clean} algorithm (H\o{}gbom 1974), using Tiny 
Tim (Krist 1995) PSFs appropriate for the positions of the galaxies on
the WFPC2 chips. The six galaxies with early-type spectra appear to
be unperturbed E and S0 galaxies (Fig.\ \ref{galaxies.plot}).
The two E+A galaxies have more
peculiar morphologies. Galaxy 1430 has a disk with
faint spiral structure, and is lopsided.
Galaxy 1583 also has a disk, and a very luminous, compact bulge. It may
be interacting with the small companion galaxy to the South.


We determined effective radii and effective surface brightnesses of the
galaxies using the 2D fitting method outlined in vDF. The values of
$r_{\rm e}$ (in arcseconds) and $\mu_{{\rm e},F814W}$ are
listed in Table 1. As noted by many
authors (e.g., JFK), the errors in $r_{\rm e}$ and $I_{\rm e}$ are
correlated, and the correlation is almost parallel to the Fundamental
Plane. Therefore, the product $r_{\rm e} I_{\rm e}^{0.83}$, which enters
the FP, can be determined to high accuracy.

\begin{small}
\begin{center}
{ {\sc TABLE 1} \\
\sc Galaxy Parameters} \\
\vspace{0.1cm}
\begin{tabular}{ccrcl}
\hline
\hline
ID & $\sigma$ & $\log r_{\rm e}$ & $\mu_{\rm e}$
& spectral type \\
\hline
        1294 &        $316 \pm 21$  &   $-0.200$  &     22.62 &
early-type \\
        1359 &        $225 \pm 19$  &   $-0.359$  &     22.33 &
early-type \\
        1405 &        $259 \pm 21$  &   $0.015$  &        23.34 &
early-type \\
        1430 &        $153 \pm 12$  &   $0.183$  &     23.88 & E+A \\
        1457 &        $210 \pm 24$  &   $-0.220$  &     22.88 &
early-type \\
        1484 &        $330 \pm 20$  &    $0.274$  &      23.84 &
early-type \\
        1567 &        $261 \pm 27$  &   $-0.283$  &      22.65 &
early-type \\
        1583 &        $212 \pm 25$  &   $-0.733$  &     20.16 & E+A \\
\hline
\end{tabular} 
\end{center}
\end{small}

\section{The Fundamental Plane and the Evolution of the $M/L$ Ratio}
\label{fp.sec}

\setcounter{figure}{2}
        
\begin{figure*}[b]
\begin{center}
\leavevmode
\hbox{%
\epsfxsize=17.5cm
\epsffile{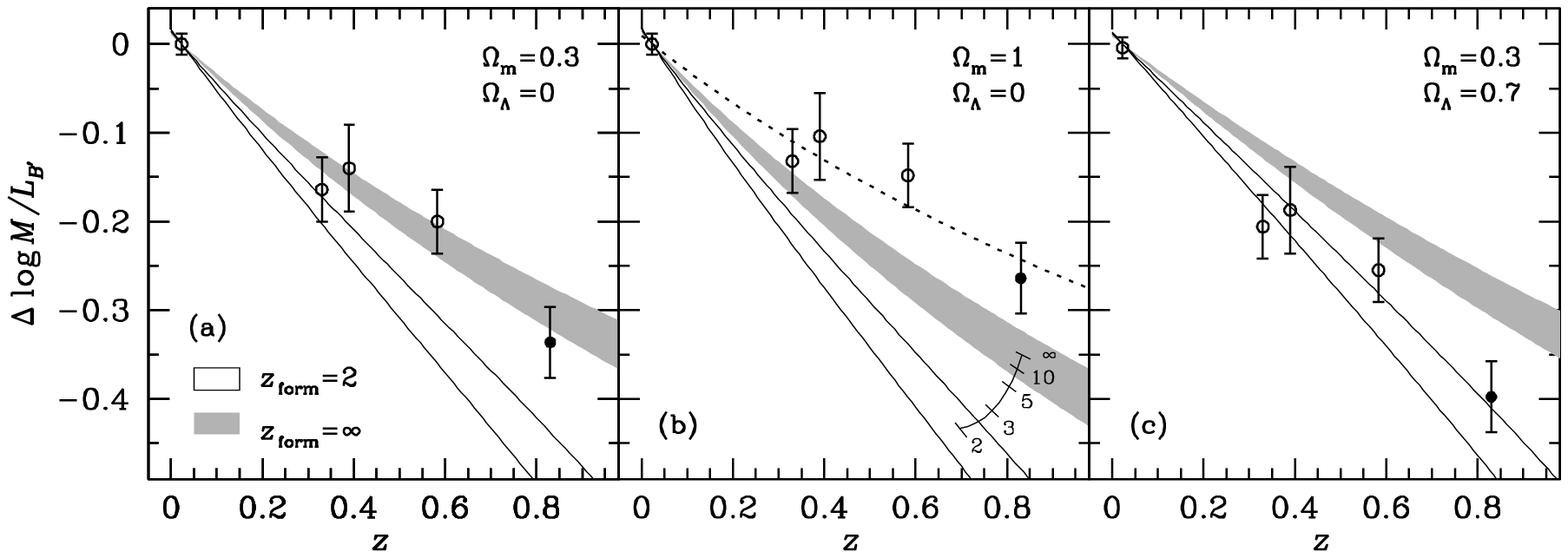}}
\figcaption{
\small Evolution of the $M/L$ ratio with redshift, for $q_0 = 0.15$ (a),
$q_0 = 0.5$ (b), and for a non-zero cosmological constant (c).
The open symbols are Coma
at $z=0.02$ (JFK), CL\,1358+62 at $z=0.33$ (KvDFIF),
CL\,0024+16 at $z=0.39$ (vDF), and MS\,2053+03 at $z=0.58$ (KvDFIF).
The solid symbol is \1054 at $z=0.83$ (this paper). Model predictions
for different formation redshifts of the stars are shown, assuming a
Salpeter IMF ($x=2.35$) and a range of metallicities. The data favor high formation
redshifts and a low value of $q_0$. The broken
line in (b) indicates a model with $z_{\rm form}=\infty$ and
a steep IMF ($x=3.35$). The conclusions are very sensitive to the IMF.
\label{ml.plot}}
\end{center}
\end{figure*}

The edge-on projection of the fundamental plane in \1054 is shown in
Fig.\ \ref{fp.plot}. The surface brightnesses ($I_{\rm e} \equiv
10^{-0.4\,\mu_{\rm e}}$) have been corrected for the $(1+z)^4$
cosmological dimming. The effective radii were converted using $q_0 =
0.15$ and $H_0 = 50$\,\kms\,Mpc$^{-1}$. The open symbols are the two E+A
galaxies. The small dots are galaxies in Coma at $z=0.023$, taken from
JFK. The line is the fit from JFK. The six galaxies with early-type
spectra define a clear FP, with low scatter ($\sim 0.045$ in
$\log\,r_{\rm e}$). The slope is similar to that of the Coma cluster. A
larger sample is needed to determine the scatter and the slope reliably.
 The offset of the FP of \1054 with respect to the FP of Coma is due to
the evolution of the $M/L$ ratio between $z=0.83$ and the present.

The E+A galaxies are over-luminous with respect
to the prediction from the FP defined by the other galaxies in \1054,
consistent with the presence of a young population. The E+As have
lower masses than the other galaxies in the sample. In the following
analysis, we will only consider the galaxies with early-type spectra.
Therefore, our conclusions may only apply to a subset of the galaxy
population. We will return to this point in Sect.\ \ref{disc.sec}.

\setcounter{figure}{1}

\vbox{
\begin{center}
\leavevmode
\hbox{%
\epsfxsize=7.5cm
\epsffile{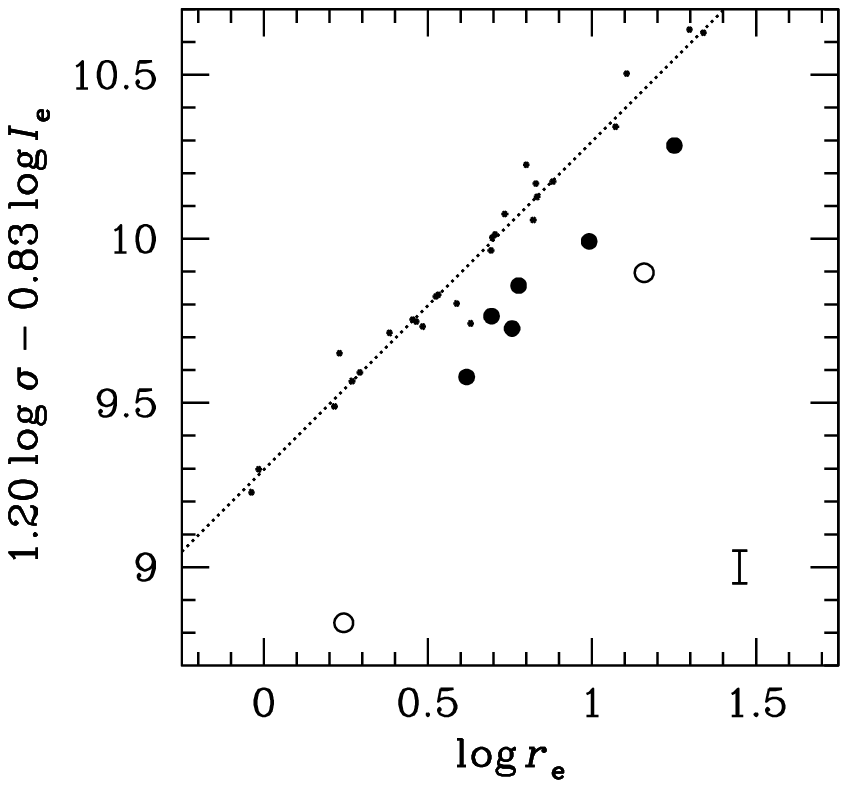}}
\figcaption{
\small Edge-on projection of the fundamental plane in \1054{} at
$z=0.83$.
The open symbols are two E+A galaxies. The error-bar indicates the
typical uncertainty. The small dots are galaxies in
Coma at $z=0.023$, from JFK. The line shows the fit from JFK.
The fundamental plane for the six galaxies with early-type
spectra is very similar to the FP of Coma.
The offset with respect to the FP of Coma is due to luminosity
evolution of the galaxies.
\label{fp.plot}}
\end{center}}

We determined the evolution of the $M/L$ ratio from the FP. In Fig.\
\ref{ml.plot}, the evolution of the $M/L$ ratio from $z=0.02$ to
$z=0.83$ is plotted.  Included in the Figure are data for Coma at
$z=0.02$ from JFK, for CL\,0024+16 at $z=0.39$ from vDF, and for
CL\,1358+62 at $z=0.33$ and MS\,2053+03 at $z=0.58$ from KvDFIF.
All data were transformed to a common rest frame band, equivalent to the
$F814W$ band in the observed frame at $z=0.83$. This band is very close
to the $B$ band redshifted to $z=0.83$, and is denoted $B_z'$. For
\1054 at $z=0.83$, $B_z' \equiv F814W + 2.5\,\log\,(1+z)$. The
transformations for the other clusters were derived in a similar way as
is described in vDF. For all clusters, the observed bands are close to
the redshifted $B$ band, and the color terms in the transformations are
small. Since the observed colors of the galaxies are used in the
transformations, the procedure is very different from a $K$-correction, 
and is independent of the spectral types of the galaxies. Using spectral
energy distributions from Pence (1976), we estimate the error induced by
the transformations is $\lesssim 0.03$ magnitudes for all spectral
types.
Because the scatter in the FP is very small, the errorbars in Fig. 3
are mainly due to systematic errors. We have assumed that the systematic
errors in the dispersions, in the photometric zeropoints, and the errors
caused by the sample selection (see vDF) can be added in quadrature.
The $M/L$ ratio evolves as $\Delta \log M/L_B
\propto (-0.40\pm 0.04)z$ if $\Omega_{\rm m}=0.3$, or
$\Delta \log M/L_B \propto (-0.31\pm 0.04)z$ if $\Omega_{\rm m}=1$.
This is consistent with other
measurements of the luminosity evolution of early-type galaxies to $z
\approx 0.8$ (e.g., Schade et al.\ 1997).
However, our measurement has a much smaller uncertainty, due to the low
scatter in the FP.

\section{Discussion}
\label{disc.sec}

The observed evolution of the $M/L$ ratio with redshift can be compared
to predictions for the evolution of single age stellar populations
formed at redshift $z_{\rm form}$. The luminosity evolution of a single
age stellar population behaves as a power law: $L \propto (t-t_{\rm
form})^{\kappa}$. The evolution depends on $\kappa$, the cosmology, and
$z_{\rm form}$ (vDF). The value of $\kappa$ depends on the
passband, the IMF, and the metallicity.  The models of Bruzual \&
Charlot (1998), Vazdekis et al.\ (1996) and Worthey (1998) give $0.86 <
\kappa_B < 1.00$ for a Salpeter (1955) IMF ($x=2.35$) and $-0.5 < {\rm [Fe/H]} < 0.5$.
Model predictions are shown in Fig.\ \ref{ml.plot}(a) for $z_{\rm form}
=2$ (open area) and $z_{\rm form}=\infty$ (shaded area), for a universe
with $\Omega_{\rm m} = 0.3$ and no cosmological constant ($q_0=0.15$).
Note that the models as well as the data points in Fig.\ \ref{ml.plot}
are independent of the value of $H_0$. The formal 95\,\% confidence
lower limit on the formation redshift is $z_{\rm form} >2.8$, for solar
metallicity, a Salpeter IMF and $q_0=0.15$. 

The slow evolution of massive galaxies in clusters not only provides a
constraint on their formation redshift, but also
on cosmological models (e.g., Bender et al.\ 1998).
This is illustrated in Fig.\ \ref{ml.plot}(b), which shows the same
models as in Fig.\ \ref{ml.plot}(a) for $\Omega_{\rm m} = 1$
($q_0=0.5$) rather than $\Omega_{\rm m}=0.3$. The models fail to fit the
data in this cosmology, and we find $\Omega_{\rm m} < 0.86$ ($95$\,\%
confidence).  The constraint on $\Omega_{\rm m}$ is
considerably stronger than the constraint derived by Bender et al.\
(1998) from the combination of the Mg$_b$ -- $\sigma$ relation and the
fundamental plane at $z=0.375$, due to the higher redshift of \1054.

These results are based on the assumption that massive early-type
galaxies in clusters have a Salpeter IMF. There is considerable
uncertainty in the slope and form of the IMF in the mass
range around $\sim 1 {\rm
M}_{\sun}$ (Scalo 1997), and there are only indirect
constraints on the IMF of massive ellipticals in clusters
(e.g., Gibson \& Matteucci 1997).
The value of $\kappa$, and hence the rate
of evolution, is quite sensitive to the logarithmic slope of the IMF:
the Worthey (1998) models give
$\Delta \kappa_B = -0.22 \Delta x$.

\setcounter{figure}{3}

\vbox{
\begin{center}
\leavevmode
\hbox{%
\epsfxsize=7.5cm
\epsffile{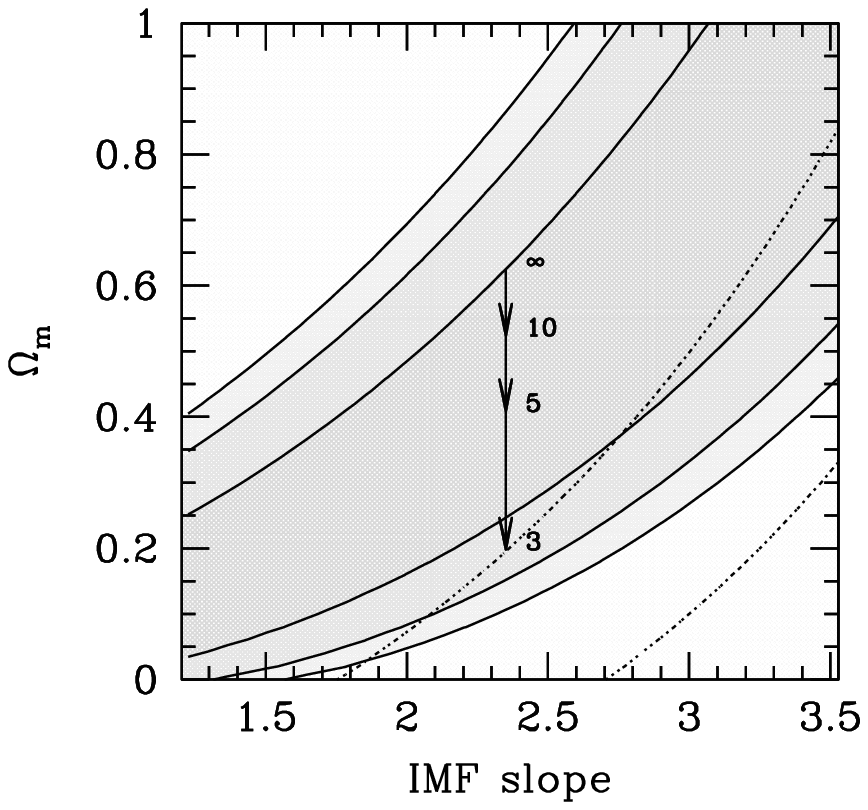}}
\figcaption{
\small Combined constraint on the slope of the IMF of massive early-type
galaxies and $\Omega_{\rm m}$, for $\Omega_{\Lambda} = 0$,
solar metallicity and $z_{\rm form} = \infty$. The solid contours
represent
confidence limits of 66\,\%, 90\,\% and 95\,\%. The arrow shows the     
effect of changing $z_{\rm form}$ to lower values. The 66\,\% confidence
limits for $z_{\rm form}=3$ are indicated with broken lines.
\label{imfomega.plot}}
\end{center}}

Since the IMF is very uncertain, we express our results as a combined
constraint on $\Omega_{\rm m}$ and the slope of the IMF in Fig.\
\ref{imfomega.plot}.
The IMF is steep if $\Omega_{\rm m} = 1$.
The 95\,\% confidence lower limit on the slope is $x>2.6$ for
$\Omega_{\rm m}=1$. As an illustration, the broken line in Fig.\
\ref{ml.plot}(b) indicates a model with $\Omega_{\rm m}=1$, $z_{\rm
form} = \infty$ and $x=3.35$. This model fits the data very well.
The case $z_{\rm form}=\infty$ is extreme;
the dependence on the formation redshift
of the constraints on $\Omega_{\rm m}$ and $x$
is indicated with an arrow
in Fig.\ \ref{imfomega.plot}.
For lower formation redshifts (or higher metallicities) the constraints
are stronger. As an example, for $z_{\rm
form} = 5$ rather than $\infty$ we find that $\Omega_{\rm m} < 0.60$ for
$x=2.35$, and $x>3.0$ for $\Omega_{\rm m}=1$.

We note that if $\Omega_{\Lambda}>0$, as is indicated by distant
supernovae (Riess et al.\ 1998),
the constraints on $z_{\rm form}$ and the slope
of the IMF are much weaker. Figure \ref{ml.plot}(c) shows the evolution
of the $M/L$ ratio if $\Omega_{\rm m}=0.3$ and $\Omega_{\Lambda}=0.7$.
The lower limit on the formation redshift is
$z_{\rm form} > 1.7$. In this cosmology, the data
provide no meaningful constraint on the slope of the IMF.

The implicit assumption in this type of study
that the set of high redshift
early-types is similar to the set of low redshift early-types (e.g.,
Kauffmann 1995, van Dokkum \& Franx 1996) is probably justified for the
high mass galaxies considered here.
It seems unlikely that a
significant fraction of the high mass galaxies in clusters at $z=0$ was
added to the sample between $z=0.83$ and $z=0$, given
the tightness of the color-magnitude relation of luminous galaxies
in intermediate $z$ clusters
(e.g., Ellis et al.\ 1997, van Dokkum et al.\ 1998),
and the low masses of the E+A galaxies (Franx 1993b, KvDFIF).
However, our conclusions may not hold for low mass galaxies: there
is good evidence that a significant fraction of the low luminosity
early-type galaxies at $z=0$ was accreted from the field at $z<1$
(e.g., Abraham et al.\ 1996, van Dokkum et al.\ 1998). 
The two E+A galaxies in our sample may be galaxies that
were recently accreted. Larger samples to fainter magnitudes are needed
to assess the fraction of galaxies with young populations in $z\approx
0.8$ clusters, and their masses.

In conclusion, the luminosity evolution of massive
early-type galaxies to $z \approx 0.8$ has now been determined to
sufficient accuracy to
place strong constraints on the epoch of formation of the galaxies, and
on cosmological models. The main uncertainty in the interpretation is
the poor understanding of the IMF in the mass range around one solar
mass. Other applications of the FP suffer from the same
uncertainty (e.g., Bender et al.\ 1998). Nevertheless, our current
measurement can be used directly to correct the evolution of the
luminosity function for the brightening of the stellar
populations with redshift. This can provide direct constraints on the
mass evolution of galaxies.

\acknowledgements{We thank M.\ Donahue for providing us with the HST
image of \1054, and G.\ Luppino for the ground based images.
The University of
Groningen and the Leids Kerkhoven-Bosscha Fonds are thanked for support.
Support from STScI grants GO07372.01-96A, GO05991.01-94A, and
AR05798.01-94A is gratefully acknowledged.}


\begin{references}
\reference{}    Abraham, R. G., Smecker-Hane,
	T. A., Hutchings, J. B.,
        Carlberg, R. G., Yee, H. K. C., Ellingson, E., Morris, S.,
        Oke, J. B., \& Rigler, M. 1996, \apj, 471, 694
\reference{}	Aragon-Salamanca, A., Ellis, R. S., Couch, W. J.,
        \& Carter, D. 1993, \mnras, 262, 764
\reference{}	Bender, R., Saglia, R. P., Ziegler, B., Belloni, P.,
	Greggio, L., Hopp, U., \& Bruzual, G. 1998, \apj, 493, 529
\reference{}	Bender, R., Ziegler, B., \& Bruzual, G. 1996, \apj, 463, L51
\reference{}	Bruzual, G., \& Charlot, S. 1998, in preparation
\reference{}	Djorgovski, S.,  \& Davis, M. 1987, \apj, 313, 59
\reference{}	Donahue, M., Voit, G. M., Gioia, I. M., Luppino, G.,
	Hughes, J. P., \& Stocke, J. T. 1998, \apj, in press
\reference{}    Dressler, A.,  \& Gunn, J. E. 1983, \apj, 270, 7
\reference{} Dressler, A., Lynden-Bell, D., Burstein, D., Davies, R. L.,
        Faber, S. M., Terlevich, R. J.,  \& Wegner, G. 1987, \apj, 313, 42
\reference{}	Ellis, R. S., Smail, I., Dressler, A., Couch, W. J.,
	Oemler, A., Jr., Butcher, H., \& Sharples, R. M. 1997, \apj,
	483, 582
\reference{}	Faber, S. M., Dressler, A., Davies, R. L., Burstein, D.,
        Lynden-Bell, D., Terlevich, R.,  Wegner, G. 1987,
        Faber, S. M., ed., Nearly Normal Galaxies. Springer, New
        York, p. 175
\reference{}	Franx, M. 1993a, \pasp, 105, 1058
\reference{}	Franx, M., 1993b, \apj, 407, L5
\reference{}	Gibson, B. K., \& Matteucci, F. 1997, \mnras, 291, L8
\reference{}	J\o{}rgensen, I., Franx, M., \& Kj\ae{}rgaard, P. 1996,
	\mnras, 280, 167
\reference{}    Kauffmann, G. 1995, \mnras, 274, 153
\reference{}    Kelson, D. D., van Dokkum, P. G., Franx, M., Illingworth,
        G. D., \& Fabricant, D. 1997, \apj, 478, L13
\reference{}	Kelson, D. D., et al. 1998, in preparation
\reference{}    Krist, J. 1995, in Astronomical Data Analysis Software
        and Systems IV, ASP Conference Series, 77, R. A. Shaw, H. E. Payne,
        and J. J. E. Hayes, eds., p. 349
\reference{}	Luppino, G. A., \& Kaiser, N. 1997, \apj, 475, 20
\reference{}	Oke, J. B., et al. 1995, \pasp, 107, 375
\reference{}	Pahre, M. 1998, Phd thesis, California Institute of
	Technology
\reference{} Pence, W. 1976, \apj, 203, 39
\reference{}	Riess, A. G., et al. 1998, \aj, in press
	(astro-ph/9805201)
\reference{}	Salpeter, E. 1955, \apj, 121, 161
\reference{}	Scalo, J. 1997, in The Stellar Initial Mass Function
	Proceedings of the 38th Herstmonceux Conference, G. Gilmore,
	I. Parry, \& S. Ryan, eds., in press (astro-ph/9712317)
\reference{}	Schade, D., Barrientos, L. F., \& Lopez-Cruz, O. 1997,
	\apj, 477, L17
\reference{}	Stanford, S. A., Eisenhardt, P. R., \& Dickinson, M.
	1998, \apj, 492, 461
\reference{}    van Dokkum, P. G., \& Franx, M. 1996, MNRAS, 281, 985  
\reference{}	van Dokkum, P. G., Franx, M., Kelson, D. D.,
	Illingworth, G. D. I., Fisher, D., \& Fabricant, D. 1998, \apj,
	500, 714
\reference{}	Vazdekis, A., Casuso, E., Peletier, R. F., \& Beckman,
	J. E., 1996, \apjs, 106, 307
\reference{}	Worthey, G. 1998, in preparation
\end{references}
\end{document}